# New Result in the Production and Decay of an Isotope, $^{278}$113, of the 113th Element

Kosuke Morita[1]*, Kouji Morimoto[1], Daiya Kaji[1], Hiromitsu Haba[1], Kazutaka Ozeki[1], Yuki Kudou[1], Takayuki Sumita[2,1], Yasuo Wakabayashi[1], Akira Yoneda[1], Kengo Tanaka[2,1], Sayaka Yamaki[3,1], Ryutaro Sakai[4,1], Takahiro Akiyama[3,1], Shin-ichi Goto[5], Hiroo Hasebe[1], Minghui Huang[1], Tianheng Huang[6], Eiji Ideguchi[7†], Yoshitaka Kasamatsu[1‡], Kenji Katori[1], Yoshiki Kariya[5], Hidetoshi Kikunaga[8], Hiroyuki Koura[9], Hisaaki Kudo[5], Akihiro Mashiko[10], Keita Mayama[10], Shin-ichi Mitsuoka[9], Toru Moriya[10], Masashi Murakami[5], Hirohumi Murayama[5], Saori Namai[10], Akira Ozawa[11], Nozomi Sato[9], Keisuke Sueki[11], Mirei Takeyama[10], Fuyuki Tokanai[10], Takayuki Yamaguchi[3], and Atsushi Yoshida[1]

[1]*RIKEN Nishina Center for Accelerator Based Science, RIKEN, Wako, Saitama 351 0198, Japan*
[2]*Faculty of Science and Technology, Tokyo University of Science, Noda, Chiba 278 8510, Japan*
[3]*Department of Physics, Saitama University, Saitama 338 8570, Japan*
[4]*Department of Chemistry, Saitama University, Saitama 338 8570, Japan*
[5]*Department of Chemistry, Niigata University, Niigata 950 2181, Japan*
[6]*Institute of Modern Physics, Chinese Academy of Science, 730000 Lanzhou, P. R. China*
[7]*Center for Nuclear Study, University of Tokyo, Wako, Saitama 351 0198, Japan*
[8]*Research Center for Electron Photon Science, Tohoku University, Sendai 982 0826, Japan*
[9]*Japan Atomic Energy Agency, Tokai, Ibaraki 319 1195, Japan*
[10]*Department of Physics, Yamagata University, Yamagata 990 8560, Japan*
[11]*University of Tsukuba, Tsukuba, Ibaraki 305 8071, Japan*

An isotope of the 113th element, i.e., $^{278}$113, was produced in a nuclear reaction with a $^{70}$Zn beam on a $^{209}$Bi target. We observed six consecutive $\alpha$ decays following the implantation of a heavy particle in nearly the same position in the semiconductor detector under an extremely low background condition. The fifth and sixth decays are fully consistent with the sequential decays of $^{262}$Db and $^{258}$Lr in both decay energies and decay times. This indicates that the present decay chain consisted of $^{278}$113, $^{274}$Rg ($Z$ 111), $^{270}$Mt ($Z$ 109), $^{266}$Bh ($Z$ 107), $^{262}$Db ($Z$ 105), and $^{258}$Lr ($Z$ 103) with firm connections. This result, together with previously reported results from 2004 and 2007, conclusively leads to the unambiguous production and identification of the isotope $^{278}$113 of the 113th element.

In the experimental program aiming at confirming the existence of an isotope of the 113th element, i.e., $^{278}$113, produced in the $^{209}$Bi + $^{70}$Zn $\rightarrow$ $^{278}$113 + $n$ reaction, we previously observed two convincing candidate events, both consisting of four consecutive $\alpha$-decays followed by a spontaneous fission (SF) immediately after the implantation of a heavy particle in the semiconductor detector under a low-background condition of typically 5 s $^{1}$.[1,2] We assigned the decay events to those originating from the primary products $^{278}$113 and the sequential decays $^{278}$113 ($\alpha_1$) $\rightarrow$ $^{274}$Rg ($Z$ = 111) ($\alpha_2$) $\rightarrow$ $^{270}$Mt ($Z$ = 109) ($\alpha_3$) $\rightarrow$ $^{266}$Bh ($Z$ = 107) ($\alpha_4$) $\rightarrow$ $^{262}$Db ($Z$ = 105) (SF).[1,2] These assignments were based on the fact that the chains connected to the known sequential $\alpha$-decays of $^{266}$Bh and $^{262}$Db as daughters of $^{278}$113.[3,4] In the continuation of our research program, we observed a new decay chain consisting of six consecutive $\alpha$-decays, i.e., $^{278}$113 ($\alpha_1$) $\rightarrow$ $^{274}$Rg ($\alpha_2$) $\rightarrow$ $^{270}$Mt ($\alpha_3$) $\rightarrow$ $^{266}$Bh ($\alpha_4$) $\rightarrow$ $^{262}$Db ($\alpha_5$) $\rightarrow$ $^{258}$Lr ($Z$ = 103) ($\alpha_6$) $\rightarrow$ $^{254}$Md ($Z$ = 101), providing an unambiguous determination of the atomic number ($Z$) and mass number ($A$) of $^{278}$113. Note that the $\alpha$-decay branching ratio of $^{262}$Db is 67%, a fact that led us to search for the $\alpha$-decay of $^{262}$Db in the present study.

This series of experiments at RIKEN Nishina Center for Accelerator Based Science was started on September 5, 2003 and tentatively terminated on August 18, 2012. The periods of the beamtime are listed in Table I together with the beam doses in each period. The net irradiation time was 553 days with a $1.35 \times 10^{20}$ beam dose in total. The first and second events obtained in 2003 2004 and 2005 2006 were published in refs. 1 and 2, respectively.

A $^{70}$Zn beam was extracted from the RIKEN Linear Accelerator. For the experiments from January 9, 2008 to July 2, 2012, a metallic bismuth layer of about 450 μg/cm$^2$ thickness was used as target. The targets were prepared by vacuum evaporation onto 30 or 60 μg/cm$^2$ carbon backing foils. The beam energy at the middle of the target layer was set to 349 MeV. The energy loss of the beam in the target was estimated to be 5.4 MeV using the range and stopping power tables;[5] the incident energies in the target ranged from 346 to 352 MeV. In the last experimental period from July 14, 2012 to August 18, 2012, where the third (present) decay chain was observed, we used thicker targets of about 780 μg/cm$^2$ thickness. The beam energy at the middle of the target layers was set to 351 MeV. The energy loss of the beam in the target was estimated to be 9.4 MeV; the beam energies ranged from 346 to 356 MeV. By increasing the target thickness, we intended to enlarge the coverage of incident beam energy so as not to miss the maximum in the excitation function of the production cross section. To avoid a possible heat problem caused by the larger energy deposit for the thicker targets, two separate foils were used as one target to reduce the energy deposit in a single target. We mounted 16 (32 for the double target system) targets on a rotating wheel 30 cm in diameter. The wheel was rotated during irradiation at 3000 4000 rpm. In the experimental period from June 12, 2012 to August 18, 2012, we replaced a Mylar vacuum-window foil of 1.1 µm thickness used to separate the gas-filled region and vacuum region for detectors with one of 0.5 µm thickness. All the other experimental conditions for the runs since 2008 were identical to those in the previous experiment.[1,2]

*E mail: morita@ribf.riken.jp
†Present address: Research Center for Nuclear Physics, Osaka University.
‡Present address: Department of Chemistry, Osaka University.

Table I. Summary of beamtime used.

| Beamtime | | Irradiation time (days) | Beam dose/sum (×10$^{19}$) | Number of observed events |
|---|---|---|---|---|
| year | month/day | | | |
| 2003 | 9/5 12/29 | 57.9 | 1.24/1.24 | 0 |
| 2004 | 7/8 8/2 | 21.9 | 0.51/1.75 | 1 |
| 2005 | 1/20 1/23 | 3.0 | 0.07/1.82 | 0 |
| 2005 | 3/20 4/22 | 27.1 | 0.71/2.53 | 1 |
| 2005 | 5/19 5/21 | 2.0 | 0.05/2.58 | 0 |
| 2005 | 8/7 8/25 | 16.1 | 0.45/3.03 | 0 |
| 2005 | 9/7 10/20 | 39.0 | 1.17/4.20 | 0 |
| 2005 | 11/25 12/15 | 19.5 | 0.63/4.83 | 0 |
| 2006 | 3/14 5/15 | 54.2 | 1.37/6.20 | 0 |
| 2008 | 1/9 3/31 | 70.9 | 2.28/8.48 | 0 |
| 2010 | 9/7 10/18 | 30.9 | 0.52/9.00 | 0 |
| 2011 | 1/22 5/22 | 89.8 | 2.01/11.01 | 0 |
| 2011 | 12/2 12/19 | 14.4 | 0.33/11.34 | 0 |
| 2012 | 1/15 2/9 | 25.0 | 0.56/11.90 | 0 |
| 2012 | 3/13 4/17 | 33.7 | 0.79/12.69 | 0 |
| 2012 | 6/12 7/2 | 15.7 | 0.25/12.94 | 0 |
| 2012 | 7/14 8/18 | 32.0 | 0.57/13.51 | 1 |
| Total | | 553 | 13.51 | 3 |

The reaction products were separated in-flight from the beam by a gas-filled recoil ion separator (GARIS) and guided into a detector box placed at the focal plane of the GARIS.[4,6] The separator was filled with helium gas at a pressure of 86 Pa. The magnetic rigidity ($B\rho$) of the GARIS to transport an evaporation residue (ER) was set to 2.09 Tm.[7] The focal plane detector system consisted of a timing detector and a silicon semiconductor detector (SSD box). The SSD box placed downstream of the timing detectors consists of five silicon detector plates. The dimensions of each detector are 60 × 60 mm$^2$. One of the silicon detectors facing the direction of incoming particles is placed at the bottom of the SSD box; it consists of 16 strip detectors (PSD). The dimensions of each strip detector are 3.75 × 60 mm$^2$. The strip detectors are position-sensitive along the longer dimension. Four other detectors (SSDs) are set to detect decaying particles from the reaction products implanted in the PSD ejected to the backward hemisphere of the PSD in coincidence with a signal from the PSD. The details of the detector are described elsewhere.[8] ERs were implanted in the PSD after passing through two timing counters. The timing signal was used for two purposes: first, to measure the time of flight (TOF) of incoming particles for the rough estimation of their mass number together with the energy signals from the PSD; second, to identify decay events originating from implanted nuclei in the PSD. The events without signals from either of the timing detectors were regarded as decay events.

The detection system was periodically checked by measuring the $\alpha$-decay lines of the ground-state transitions of the transfer-reaction products, $^{211}$Po (7.4503 ± 0.0005 MeV), and $^{213}$Rn (8.088 ± 0.008 MeV), setting the $B\rho$ of GARIS at 1.67 Tm, and keeping all the other conditions the same as those for the actual measurement. The energy calibration of the detectors for decay $\alpha$-particles was performed simultaneously. Note that, because the $\alpha$-decays used for the energy calibration were ground-state transitions, no $\gamma$-ray was emitted with the $\alpha$-decays.

The $\alpha_1$-, $\alpha_2$-, $\alpha_4$-, $\alpha_5$-, and $\alpha_6$-particles were detected by only the PSD. The energy resolution measured using only the PSD was 55 keV in full width at half maximum (FWHM). For $\alpha_3$, the decaying $\alpha$-particles were ejected from the PSD and implanted to the SSD. Therefore, the energies of the decaying $\alpha$-particles were measured partly by the PSD and those of the residual ones by the SSD. The resolution in the case that the energy of a decaying particle was measured by both the PSD and the SSD was 73 keV in FWHM. The position resolution of the PSD for the ER and $\alpha$-measurements was measured to be 0.58 mm in FWHM.

The decay chain observed in this work consisted of six consecutive $\alpha$-decays after the implantation of an ER in the focal plane detector of the GARIS. The ER and all of the $\alpha$-particles were detected at strip #7 of PSD. The $\alpha$-energies, time intervals from the previous decay (or implantation of ER), and positions in a strip of PSD of the decays are listed in the third column of Table II together with those observed in previous runs (first and second columns). In the third $\alpha$-decay of the second chain, we consider that only a portion of the $\alpha$-energy was measured owing to the limited solid angle coverage (85%) of the detector box. In Table II, the times indicated in the cells for ER in nanosecond units are the TOF measured along the 295 mm flight path just before implantation into the focal plane detector. The mean lifetimes obtained from the three decay chains are also listed in the last column in Table II with 1$\sigma$ statistical errors for $\alpha_1$, $\alpha_2$, $\alpha_3$, $\alpha_4$, and SF/$\alpha_5$. The value for $\alpha_6$ is simply the observed one for chain 3 with a 1$\sigma$ statistical error for one event.

For the ER event, the TOF between the timing counters was 42.6 ± 0.5 ns, which was somewhat shorter than those of the first and second events (44.6 ± 0.5 and 45.7 ± 0.5 ns) because of a finite difference of the kinetic energies of ER, as discussed below. The implantation energies observed for the first, second, and third events were 36.8, 36.5, and 41.9 MeV, respectively. The kinetic energies of ER for the first, second, and third events were calculated using the TOFs along the 295 mm path length to be 62.9, 59.9, and 69.0 MeV, respectively, assuming that the atomic mass number of the ER is 278. By taking into account the reaction kinematics, and energy losses in the target, in the filling gas, in the foil separating the gas and detection regions, and in the foil of the first TOF counter using the energy loss table,[5] the kinetic energies of ERs just after the first foil of the timing counter are calculated to be 62 MeV for the first and second events, and 66 MeV for the third event. These values are in good agreement considering the uncertainty of the energy losses of ERs in the target due to the depths of the reaction points in the target that we could not determine. The differences between the observed implantation energies for the three events, i.e., 36.8, 36.5, and 41.9 MeV, and the values calculated using the TOFs, i.e., 62.9, 59.9, and 69.0 MeV, are well described by the energy losses in the foil of the second timing counter and pulse height defects, the so-called plasma effect, and nuclear stopping in silicon PSD,[9)] for ions with a large atomic number.

**Table II.** Summary of observed events. Numbers in parentheses denote energy resolutions in FWHM. $\Delta T$: time difference from the preceding decay (or the implantation of ER). Position: measured from the bottom of the silicon strip in which ER was implanted. The time indicated in a cell for ER is the TOF along the 295 mm flight path measured using the timing counters.

| | Chain 1[1)] | Chain 2[2)] | Chain 3 (present) | |
|---|---|---|---|---|
| | $E$ (MeV) | $E$ (MeV) | $E$ (MeV) | Assignment |
| | $\Delta T$ | $\Delta T$ | $\Delta T$ | mean lifetime |
| | Position | Position | Position | |
| ER | 36.75 | 36.47 | 41.91 | |
| | 44.6 ns (TOF) | 45.7 ns (TOF) | 42.6 ns (TOF) | |
| | 30.3 mm | 30.1 mm | 4.9 mm | |
| $\alpha_1$ | 11.68 (0.04) | 11.52 (0.04) | 11.82 (0.06) | $^{278}$113 |
| | 0.344 ms | 4.93 ms | 0.667 ms | $2.0^{+2.7}_{-0.7}$ ms |
| | 30.5 mm | 30.2 mm | 4.4 mm | |
| $\alpha_2$ | 11.15 (0.07) | 11.31 (0.07) | 10.65 (0.06) | $^{274}$Rg |
| | 9.26 ms | 34.3 ms | 9.97 ms | $18^{+24}_{-7}$ ms |
| | 30.4 mm | 29.6 mm | 4.8 mm | |
| $\alpha_3$ | 10.03 (0.07) | 2.32 (escape) | 10.26 (0.07) | $^{270}$Mt |
| | 7.16 ms | 1.63 s | 444 ms | $0.69^{+0.95}_{-0.26}$ s |
| | 29.8 mm | 29.5 mm | 5.1 mm | |
| $\alpha_4$ | 9.08 (0.04) | 9.77 (0.04) | 9.39 (0.06) | $^{266}$Bh |
| | 2.47 s | 1.31 s | 5.26 s | $3.0^{+4.2}_{-1.1}$ s |
| | 30.9 mm | 29.7 mm | 4.9 mm | |
| SF/$\alpha_5$ | 204 (SF) | 192 (SF) | 8.63 (0.06) | $^{262}$Db |
| | 40.9 s | 0.787 s | 126 s | $56^{+77}_{-21}$ s |
| | 30.3 mm | 30.5 mm | 4.5 mm | |
| $\alpha_6$ | | | 8.66 (0.06) | $^{258}$Lr |
| | | | 3.78 s | $3.8^{+18}_{-1.7}$ s |
| | | | 4.7 mm | |

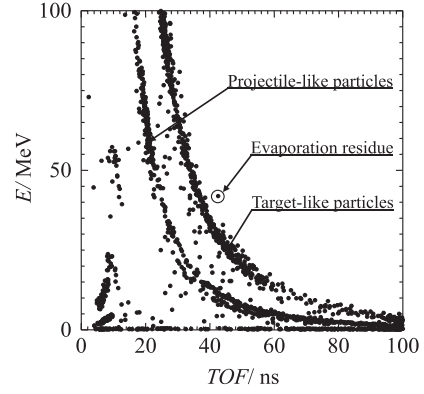

**Fig. 1.** Two dimensional plot of energy measured by PSD vs TOF measured using timing counters for the run (time period = 10.8 h), in which the decay chain was observed. Events detected by only strip #7 of PSD are shown. A point corresponding to the implantation event is shown by a circle with an arrow. Loci corresponding to projectile like particles ($A \approx 70$) and target like particles ($A \approx 209$) are also shown.

Figure 1 shows a scatter plot of the energy and TOF for the run in which the decay chain was observed. In the figure, only events detected by strip #7 of the PSD are plotted. The point corresponding to the implantation event is indicated by a circle with an arrow. Loci corresponding to projectile-like particles ($A \approx 70$) and target-like particles ($A \approx 209$) (target recoil and transfer reaction products) are seen in the figure. The implantation event is well separated from the locus of the target-like particles that may contribute to the background of the measurement. The figure indicates that the mass of the implanted particle is higher than that of the target-like particles. The mass number of ER was roughly estimated to be $291 \pm 15$ using a method similar to that described in ref. 8.

The probabilities of an accidental coincidence between the implantation of ER and individual decays were estimated as follows. The counting rate of decay events at strip #7, which yielded no TOF signal for a decay energy greater than 8 MeV, was $2.6 \times 10^{-3}$ s$^{-1}$. The decay events were uniformly distributed along the strip, and a position window for decay identification was set to 0.6 mm, which is the measured position resolution in FWHM, over the 60-mm-long strip. Thus, the effective counting rate was $2.6 \times 10^{-5}$ s$^{-1}$. With time differences of 0.667 ms for $\alpha_1$, 10.6 ms for $\alpha_2$, 455 ms for $\alpha_3$, 5.7 s for $\alpha_4$, 131 s for $\alpha_5$, and 135 s for $\alpha_6$, the probabilities of the accidental coincidence were evaluated to be $1.7 \times 10^{-8}$, $2.8 \times 10^{-7}$, $1.2 \times 10^{-5}$, $1.5 \times 10^{-4}$, $3.5 \times 10^{-3}$, and $3.5 \times 10^{-3}$, for ER-$\alpha_1$, ER-$\alpha_2$, ER-$\alpha_3$, ER-$\alpha_4$, ER-$\alpha_5$, and ER-$\alpha_6$, respectively. The probability of producing all six of these decays accidentally is the product of these numbers, i.e., $1.1 \times 10^{-28}$, which excludes the possibility of an accidental coincidence.

The $\alpha$-decay mode was first observed for the fifth decay in the third (present) chain, while the modes observed in the first and second decay chains were the SF decay. The observed $\alpha$-energy of $E_\alpha = 8.63 \pm 0.06$ MeV is in good agreement with the adopted values for $^{262}$Db: $E_\alpha = 8.450 \pm 0.020$ MeV ($I_\alpha = 75\%$), $8.530 \pm 0.020$ MeV (16%), and $8.670 \pm 0.020$ MeV (9%).[10)] The mean life obtained from the three decays (1 $\alpha$-decay and 2 SFs) is $56^{+77}_{-21}$ s which corresponds to the half-life of $T_{1/2} = 39^{+53}_{-14}$ s. The obtained $T_{1/2}$ value is also in good agreement with the adopted value of $T_{1/2} = 34 \pm 4$ s.[10)]

In the present decay chain, the sixth $\alpha$-decay ($\alpha_6$) of $E_\alpha = 8.66 \pm 0.06$ MeV was also observed with a decay time of 3.78 s, which corresponds to $T_{1/2} = 2.6^{+12}_{-1.1}$ s. These decay properties are in good agreement with the adopted properties for the $\alpha$-decay daughter of $^{262}$Db, $^{258}$Lr: $8.565 \pm 0.025$ MeV (20%), $8.595 \pm 0.010$ MeV (46%), $8.621 \pm 0.010$ MeV (25%), and $8.654 \pm 0.010$ MeV (9%), $T_{1/2} = 3.92^{+0.35}_{-0.42}$ s and $b_\alpha = 97.5\%$.[10)]

The decay time distributions of individual generations in the three decay chains are shown in Fig. 2, where the logarithm of the decay time is taken as the abscissa. The distribution of a simple one-component exponential decay is expressed by a Gaussian-like universal curve whose peak position corresponds to the mean life with a tail on the shorter-decay-time side.[11)] It was found that the decay time of each generation closely follows a single universal curve, as shown in Fig. 2. This indicates that the mean lifetime of each generation has a single component; the observed decays originated from nuclides with the same mean life.

The $\alpha$-decay energies of $^{278}$113, $^{274}$Rg, $^{270}$Mt, and $^{266}$Bh show some discrepancies among the three chains. However,

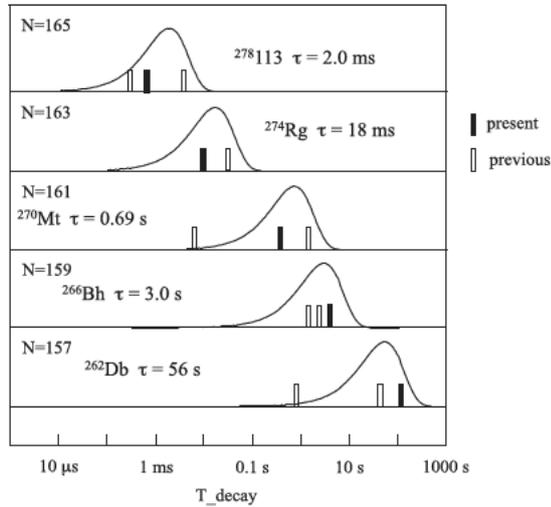
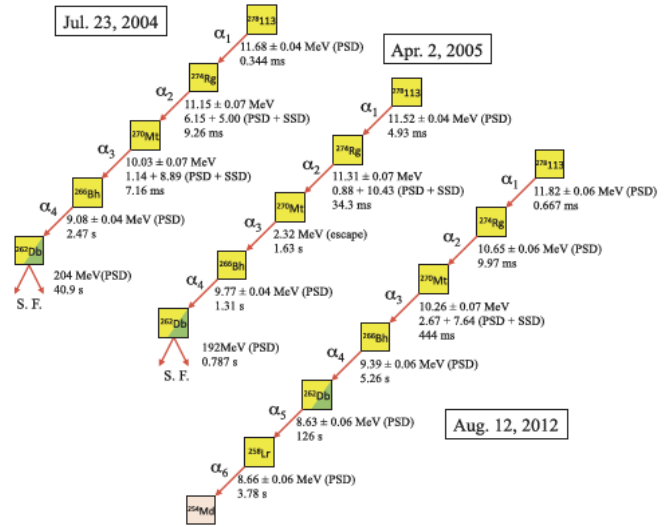

Fig. 2. Decay time (T_decay) distributions of the decay family originating from $^{278}113$ are indicated. The logarithm of the decay times is taken as the abscissa. Mean lifetimes $\tau$ determined from three decay chains are shown together with the symbols of the nuclides. Curves in the graphs correspond to single exponential decay curves with $\tau$. Neutron numbers of the nuclides are also indicated in the graphs.

the observation in the rather widely distributed decay energies of the decay chains of odd-odd nuclei, for example, starting from $^{272}$Rg,[8] is a natural feature due to the decays to the many excited states in their daughters, and due to the possible summing effect of $\alpha$-energy with a conversion electron or $\gamma$-ray energy, which is emitted simultaneously with $\alpha$-decay inside the detector.

Therefore, we could conclusively assign the sixth decay to that of $^{258}$Lr. Consequently, we could unambiguously assign the third decay chain to $^{278}113 \rightarrow {}^{274}$Rg $\rightarrow {}^{270}$Mt $\rightarrow {}^{266}$Bh $\rightarrow {}^{262}$Db $\rightarrow {}^{258}$Lr $\rightarrow {}^{254}$Md.

The isotopes $^{254}$Md/$^{254m}$Md are known to decay dominantly by electron capture (EC) ($b_\varepsilon \approx 100\%$) with half-lives $T_{1/2} = 10 \pm 3$ min and $28 \pm 8$ min, respectively.[12] We did not observe the corresponding event following the assigned decay of $^{258}$Lr, because the energy deposit of X-rays (the highest energy of characteristic X-rays of $^{254}$Fm is 142 keV)[13] associated with the EC decay is smaller than the energy threshold of the focal plane detector, i.e., 800 keV. The isotope $^{254}$Fm, which is the daughter of $^{254}$Md, is known to decay dominantly by $\alpha$-emission ($b_\alpha = 99.94\%$) with a half-life $T_{1/2} = 3.240 \pm 0.002$ h.[12] The adopted decay energies and relative intensities are $6.898 \pm 0.003$ MeV (0.0066%), $7.050 \pm 0.002$ MeV (0.82%), $7.150 \pm 0.002$ MeV (14.2%), and $7.192 \pm 0.002$ MeV (84.9%).[12] We observed two candidate events corresponding to the decay of $^{254}$Fm. One ($\alpha_{7\_1}$) was observed 3.96 h after the decay of $^{258}$Lr with a decay energy and a position in the detector of 7.26 (0.07) MeV and 5.2 mm, respectively. The other ($\alpha_{7\_2}$) was observed 6.42 h after the decay of $^{258}$Lr with a decay energy and a position of 7.18 (0.06) MeV and 5.1 mm, respectively. Using the counting rate of the decay event mentioned above, but for a decay energy greater than 7 MeV, i.e., $3.0 \times 10^{-3}$ s$^{-1}$, the probabilities of an accidental coincidence between the implantation of ER and the observed decays, $\alpha_{7\_1}$ and $\alpha_{7\_2}$, were estimated to be 0.43 and 0.70, respectively. We consider that one of these was the possible decay of $^{254}$Fm.

Fig. 3. (Color) Observed decay chain in the present work together with previously observed chains.[1,2]

The decay daughter of $^{254}$Fm, $^{250}$Cf, is known to decay dominantly by $\alpha$-emission ($b_\alpha = 99.92\%$) with a half-life $T_{1/2} = 13.08 \pm 0.09$ years.[12] We have not yet observed the corresponding event.

The observed decay chains are shown in Fig. 3 together with the previous two. The total beam dose was $1.35 \times 10^{20}$. Combining all three events, the production cross section of $^{278}113$ was determined to be $22^{+20}_{-13}$ fb (fb = $10^{-39}$ cm$^2$) with a $1\sigma$ error. The error includes only a statistical one. To deduce the cross section, the values of the transmission of the GARIS and the effective target thickness used were 0.8 and 450 $\mu$g/cm$^2$, respectively.

In conclusion, the isotope of the 113th element, i.e., $^{278}113$, was produced in the $^{209}$Bi($^{70}$Zn,n)$^{278}113$ reaction and was unambiguously identified by the firm connection to the well-known daughter nuclides $^{266}$Bh, $^{262}$Db, and $^{258}$Lr.

**Acknowledgments** The experiment was performed at the RI Beam Factory operated by RIKEN Nishina Center and CNS, University of Tokyo. We would like to thank the accelerator staff for their excellent operation and assistance during our extremely long experiment. Special thanks are due to all members of RIKEN Headquarter headed by Dr. Ryoji Noyori for their support, which allowed us to perform the experiment during the severe limitation of electric power owing to the serious accident at the Fukushima nuclear power plant. This article is dedicated to all the people who were lost or injured in the devastating earthquake and tsunami of March 11, 2011, that occurred in the northeast area of Japan. This research was partly supported by a Grant-in-Aid for Specially Promoted Research, 19002005, 2007, from the Ministry of Education, Culture, Sports, Science and Technology, Japan. This work was also partly supported by the RIKEN Strategic Program for R&D.